\documentstyle[preprint,aps,psfig]{revtex}

\begin{document}
\tighten
\preprint{\vbox{\hbox{IFP-769-UNC}
\hbox{hep-ph/9903479} 
\hbox{March 1999}}
}

\title{Unification with Enlarged Kaluza-Klein Dimensions}
\author{Paul H. Frampton and Andrija Ra\v{s}in}

\address{{\it Department of Physics and Astronomy}}

\address{{\it University of North Carolina, Chapel Hill, NC 27599-3255}}

\maketitle

\begin{abstract}
In minimal theories with extra spatial dimensions at scales $\mu_0$ 
much lower than the conventional GUT scale, unification 
can give too-large 
predictions for $\alpha_3(M_Z)$ 
given $\alpha_1(M_Z)$ and $\alpha_2(M_Z)$ as empirical input.
We systematically study the effects of adding extra states
above the compactification scale 
on running of the gauge couplings 
and find several simple examples that give 
unification where all $\alpha_i(M_Z)$ are consistent with low-energy data.
We study both the supersymmetric and nonsupersymmetric
unification.

\end{abstract} 

\newpage

{\it Introduction.} \hspace{0.5cm} Theories with extra space-time 
dimensions arise naturally from superstrings. In particular, extra
dimensions may appear at lower scales than the Planck scale, for example
at the grand unified scale\cite{w96}. One can entertain the possibility
that the compactification scale may lie much lower
\cite{a90,l96,add98,adpq98,ddg98,ka98,biq98,st98,b98,ddgalpha,gr98,c99,add982},
even allowing the string scale at 1TeV where one
could hope for experimental signatures in the foreseeable future.

For the string scale lying within the TeV region striking signatures
are expected depending on the details of a specific model (such
as the sizes of the compactified dimensions inside and transverse to the
$p$-branes), ranging from
modification of the Newton's law of gravitation at submillimeter
distances\cite{add982}, collider signatures\cite{mpp98}, large muon
anomalous magnetic moment\cite{mg99}, deviations of electroweak 
observables\cite{mp99} and other phenomena\cite{oth}.

Such theories with compactification scale lower 
than the usual GUT scale of $10^{16}$ GeV also
strongly impact effects that are associated with scales higher than the
weak scale such as proton decay, neutrino masses, flavor changing neutral 
currents, etc. 

If one believes in gauge coupling unification, one can
naturally ask how well such couplings unify. Once the gauge couplings
cross the compactification scale their running changes on a $p$-brane
for $p>3$ from logarithmic to power-law\cite{tt88}. In fact, at
energies above the compactification scale $\mu_0$ we can think of an
effective higher dimensional theory.
For example, one can show that in the
case with one ${\cal N}=2$ Higgs hypermultiplet above $\mu_0$
\cite{ddg98}
the beta functions in extra dimensions were such that the couplings
still approximately unify albeit at the scale very near the
compactification scale, because of the power-law running in extra
dimensions. The unification is only approximate: given $\alpha_{1,2}(M_Z)$
the prediction for $\alpha_3(M_Z)$ depends on the number of extra dimensions
and the compactification scale, but it is typically much higher than the
experimental value as soon as the compactification scale is lowered
below $10^{16}$ GeV. For example, for one extra dimension, 
$\alpha_3(M_Z) \approx 0.13$ at $\mu_0 = 10^{14}$ GeV
and goes up to $\alpha_3(M_Z) \approx 0.16$ at $\mu_0 = 10^5$ GeV.
Their analysis was further refined at the two-loop
level in \cite{gr98}, where it was shown that $\alpha_3(M_Z)$
was further increased. As we will show the situation is even worse
in the minimal scenario with two Higgs hypermultiplets
above $\mu_0$\footnote{We call "minimal" the scenario where each
Higgs supermultiplet has its own tower of Kaluza-Klein states,
which above $\mu_0$ means two ${\cal N}=2$
hypermultiplets. However, depending on the details of the theory,
one can allow for 0,1 or 2 Higgs hypermultiplets and we discuss all
scenarios later in the text.}

One can adopt two attitudes to this situation. The first is simply to abandon
unification. After all, $SU(5)$ or $SO(10)$ at scales lower than
$10^{16}$GeV would simply be disastrous for the proton decay rate.
Indeed proton decay is a delicate issue for these enlarged Kaluza-Klein 
dimensional theories.

But the second more constructive attitude
is to try to fix unification with conventional means
like adding extra states or scales, while disallowing proton decay
on symmetry grounds. Unification into groups like $SU(3)^3$
or Pati-Salam, would avoid proton decay. Other
solutions to avoid proton decay that merit further investigation
include compactification defined on 
a $Z_2$ orbifold, where interactions that lead to proton
decay vanish at the orbifold fixed points\cite{ddg98} or
gauging and breaking an additional U(1) symmetry on a distant
brane\cite{add98,ka98,st98}.

In this paper we assume that the proton decay problem is
dealt in some satisfactory manner and we concentrate on the
issue of precise gauge unification. We systematically
study for which additional field representations unification
can be achieved. We assume that extra representations appear
at the compactification scale $\mu_0$. We do not add
any additional states at $M_Z$. 
For the usual case without any extra dimensions couplings 
do unify in the MSSM to a high degree\cite{ms81}, 
with the predicted strong coupling value $\alpha_3(M_Z)\approx 0.126$
somewhat higher than the world average 
$\alpha_3(M_Z)=0.119 \pm 0.002$.
The non-supersymmetric running with extra low lying states has been 
extensively studied since the early days\cite{my92}. 
One can also include extra matter at some new intermediate scale
but we do not consider this possibility in this paper.

We study first the supersymmetric
case and search minimal $SU(3)\times SU(2)\times U(1)$
irreducible representations
(minimal in the sense that they are contained in (broken)
irreducible representations of $SU(5)$ up to dimension {\bf 75}). 
We find several simple representations which predict $\alpha_3(M_Z)$
within experimental
limits. Since we are giving up the proton decay constraints
and we are adding extra states, it is well justified to ask
what happens in the nonsupersymmetric case as well. This
case is taken up before we formulate our conclusions.

\bigskip
\bigskip

{\it Supersymmetric Unification.}  \hspace{0.5cm} First we derive a simple condition on the
beta functions at one-loop level that such extra states must
satisfy. We then among those candidates, run the full two-loop
analysis similar to Ref. \cite{gr98}.

The evolution equations are

\begin{equation} 
\alpha^{-1}_i(M_z) = \alpha_i^{-1}(\mu) + { b_i \over
{2\pi}} {\rm ln} \left[ {\mu \over M_z} \right]
, \,\, M_Z < \mu <
\mu_0 \,\, ,
\label{ddg1}
\end{equation}
below the compactification scale $\mu_0$. Above $\mu_0$ in a theory
with $4+\delta$ dimensions we have
\footnote{This one-loop equation is actually an approximation to an
analytical expression that involves an integral over a Jacobi theta
function\cite{ddg98}. However, as shown in \cite{gr98}, which analysis we
follow later, this approximation does not affect the two-loop results
for low number of extra dimensions ($d = 1-3$) that we consider here.}

\begin{equation} 
\alpha^{-1}_i(\mu_0) = \alpha_i^{-1}(\mu) + 
{ {(b_i - \tilde{b}_i)} \over {2\pi}} 
{\rm ln} \left[ {\mu \over \mu_0} \right]
+{{\tilde{b}_iX_\delta} \over {2\pi\delta}}\left[
\left({\mu \over \mu_0}\right)^\delta - 1 \right], \,\, \mu_0 < 
\mu \, \, ,
\label{ddg2}
\end{equation}
where $b_i=(33/5,1,-3)$ are the MSSM beta functions, 
$\tilde{b}_i$ are the beta functions corresponding to the 
excited Kaluza-Klein states and 
$X_\delta = {{\pi^{\delta \over 2}} \over 
{\Gamma(1+{\delta \over 2})}}$. 

{\bf 1. Case with two Higgs hypermultiplets.}
In the scenario where both MSSM Higgses have their corresponding massive
Kaluza-Klein towers
\begin{equation}
\tilde{b}^{min}_i=(6/5,-2,-6)+\eta(4,4,4)
\label{minddg}
\end{equation}
come from  contributions from excited Kaluza-Klein states
of the standard model fields, including
the gauge bosons, Higgs doublets and
$\eta$ chiral generations with $\eta=0-3$. 
Allowing for $\eta \neq 3$ means that we allow also some generations
to have no excited states. One may as well
consider the possibility that some of the other standard 
model states have no Kaluza-Klein excitations. Unification with
only $SU(3) \times U(1)$ or $SU(2) \times U(1)$ or just $U(1)$ in the
bulk was considered in \cite{c99}. We will later consider also
the possibility of putting different number of Higgs
doublet hypermultiplets.

The procedure is as follows.
We will consider the addition of 
extra states above $\mu_0$ which will change the beta functions to
\begin{equation}
\tilde{b}_i = \tilde{b}^{min}_i+\Delta\tilde{b}_i \,\, .
\label{deltaddg}
\end{equation}
We proceed by first considering the one-loop part
to find an analytic condition on $\Delta\tilde{b}_i$.
For the candidates that most closely satisfy the approximate one-loop
condition we do the two-loop analysis to find the predictions
for $\alpha_3(M_z)$.

Now we require unification at a scale $\Lambda > \mu_0$.
Combining equations (\ref{ddg1}) and (\ref{ddg2}) and eliminating
$\alpha^{-1}(\Lambda)$ we get at one-loop
\begin{equation}
\alpha^{-1}_i(M_z) - \alpha^{-1}_j(M_z) = 
{{(b_i-b_j)} \over {2\pi}}
\left\{
{\rm ln} {\Lambda \over M_z} - 
B_{ij} {\rm ln} {\Lambda \over \mu_0} +
B_{ij} { {X_\delta} \over {\delta}} 
\left[\left({ \Lambda \over \mu_0}\right)^\delta - 1 \right]
\right\} \,\, ,
\end{equation}
where
\begin{equation}
B_{ij} \equiv {{\tilde{b}_i - \tilde{b}_j} \over {b_i - b_j}}
\, \, .
\end{equation}
Notice that this formula actually tells us that the couplings would 
unify to the precision of the MSSM if $B_{ij}$ were not dependent
on $i$ and $j$, so that the curly bracket on the rhs would be
a constant with the size of order ln${{10^{16} \rm GeV} \over M_Z}$.
Thus as a first approximation of unification we may consider the condition
\begin{equation}
 {B_{12} \over B_{13}} = {B_{13} \over B_{23}} = 
{B_{12} \over B_{23}} = 1 \, \, .
\label{mssmcond}
\end{equation}
For the minimal case (\ref{minddg}) we have
\begin{equation}
 {B^{min}_{12} \over B^{min}_{13}} = 0.76 \, \, , 
 {B^{min}_{13} \over B^{min}_{23}} = 0.75 \, \, , 
 {B^{min}_{12} \over B^{min}_{23}} = 0.57\, \, , 
\end{equation}
which signals non-unification. Indeed, as shown in Table 2, 
$\alpha_3(M_Z)$ gets {\it higher} as $\mu_0$ is lowered
and is already 0.16 at $\mu_0=10^{14}$ GeV.

Let us now assume that we add extra fields above $\mu_0$
so that the beta functions get changed by $\Delta\tilde{b}_i$
as in (\ref{deltaddg}).
Applying the condition (\ref{mssmcond}) translates into
\begin{equation}
5\Delta\tilde{b}_1 - 12\Delta\tilde{b}_2 + 7\Delta\tilde{b}_3 = 12
\, \, ,
\label{thecondn}
\end{equation}
with the right hand side being non-zero signaling the 
nonunification in the minimal case $\Delta\tilde{b}_i=0$.

In Table 1, we list all the $SU(3)) \times SU(2) \times U(1)$ states that
are contained in
the lowest lying $SU(5)$ multiplets (up to $\underline{\bf 75}$)
that have the left hand side of the above equation between
-60 and 60. Notice that above $\mu_0$ states are in complete
${\cal N}=2$ hypermultiplets which consist of two chiral multiplets,
and that in addition we take two hypermultiplets with opposite charges in
order to cancel anomalies.

From Table 1 we see that some simple examples of states that 
perfectly rectify the one-loop gauge unification include
\begin{eqnarray}
\begin{array}{c}
 (1,1)_{\pm 1} \, \, {\rm  hypermultiplets}\cite{ka98} 
\, \, ; \, \, {\rm or}\\ 
(1,3)_{\pm 2/3} \,\, {\rm  and } \,\,(2,1)_{\pm 1/2} \,\, 
{\rm hypermultiplets}
\, \, ; \, \, {\rm or}\\ 
(2,1)_{\pm 3/2} \,\, {\rm  and } \,\,(2,1)_{\pm 1/2} \,\, 
{\rm hypermultiplets}
\, \, ; \, \, {\rm or}\\ 
{\rm two  } \,\, (2,3)_{\pm 5/6} \,\, {\rm  hypermultiplets}
\, \, ; \, \, {\rm or}\\ 
(2,3)_{\pm 7/6} \,\, {\rm  and } \,\,(2,3)_{\pm 1/6} \,\, 
{\rm hypermultiplets}
\,\,.
\end{array}
\label{candidates}
\end{eqnarray}
One still has to check for consistency conditions
such as whether the couplings remain perturbative.
We have performed a numerical, two-loop analysis
similarly to Ref. \cite{gr98}\footnote{The two-loop analysis of
\cite{gr98}
actually includes only the two-loop evolution below $\mu_0$, and does
not include the two-loop power-law behavior.}.
for the
states with 
$5\Delta\tilde{b}_1 - 12\Delta\tilde{b}_2 + 7\Delta\tilde{b}_3$
equal to or close to 12.
The results are shown in Table 3.

From Tables 3a-3d, we note as expected that the combinations in 
(\ref{candidates}) all have $\alpha_3(M_Z)$ within
the MSSM value (or even closer to the
world average $\alpha_3(M_Z)=0.119\pm0.002$) for 
basically all values of $\mu_0$ between the TeV scale and
$10^{16}$ GeV. However, for the last two candidates
with extra $(2,3)$ multiplets perturbativity
breaks down for lower $\mu_0$\footnote{Note that perturbativity
here is actually the requirement that not just the
coupling strength be less than 1, but 
the product of the inverse coupling strength and the 
number of KK states (running in the loops).}.

In Tables 3e-3f we have also shown predictions for
$\alpha_3(M_Z)$ for some states which  are close to
the condition in (\ref{thecondn}). Now of course, the
prediction for $\alpha_3(M_Z)$ varies with $\mu_0$.
We note an interesting property. The states that have 
$5\Delta\tilde{b}_1 - 12\Delta\tilde{b}_2 + 7\Delta\tilde{b}_3$
higher than 12 have lower $\alpha_3(M_Z)$ as $\mu_0$ is lowered
and can actually have values compatible with experiments down to 
intermediate scales. In the opposite case, when
$5\Delta\tilde{b}_1 - 12\Delta\tilde{b}_2 + 7\Delta\tilde{b}_3$
is lower than $12$, $\alpha_3(M_Z)$ gets larger with lower $\mu_0$.

One could object to having states at intermediate scales
which have exotic quantum numbers. However, such states are
easily found in multiplets of grand unified theories, and indeed, 
in ordinary supersymmetric theories with more mass scales, it is easy 
to get extra light exotic states\cite{ams97}. In any case, a 
complete unification program should have a complete theory, 
for which one computes the mass spectrum of states and performs the
runnings. Analysis such as presented in this paper should be a 
guiding point in search for a more complete theory.

{\bf 2. Cases with one or zero Higgs hypermultiplets.}
Let us also remark on the possibility that above $\mu_0$
one adds or drops the Kaluza-Klein towers
of Standard Model states. Of course, dropping or adding complete
generations 
has no effect on unification except on the size of the
coupling at the unification scale.
The case where some of the gauge bosons are left out was studied in
\cite{c99}. If we assume that there is only one Higgs
${\cal N}=2$ hypermultiplet above $\mu_0$
in which case\cite{ddg98}
\begin{equation}
\tilde{b}^{min,1higgshm}_i=(3/5,-3,-6)+\eta(4,4,4)
\end{equation}
and the unification condition corresponding to (\ref{thecondn}) is
\begin{equation}
5\Delta\tilde{b}_1 - 12\Delta\tilde{b}_2 + 7\Delta\tilde{b}_3 = 3
\, \, .
\label{diffcon}
\end{equation}
For this case we have
\begin{equation}
 {B^{min}_{12} \over B^{min}_{13}} = 0.94 \, \, , 
 {B^{min}_{13} \over B^{min}_{23}} = 0.92 \, \, , 
 {B^{min}_{12} \over B^{min}_{23}} = 0.86 \, \, , 
\end{equation}
which signals approximate unification\cite{ddg98}. However as
noted in \cite{ddgalpha,gr98,c99}, the approximate unification
translates into $\alpha_3(M_Z)$ as high as $0.116-0.117$ for
lower values of $\mu_0$. Notice that again this is the same phenomenon
as described above, that when the 
$5\Delta\tilde{b}_1 - 12\Delta\tilde{b}_2 + 7\Delta\tilde{b}_3$ 
is less than 3, $\alpha_3(M_Z)$ grows as $\mu_0$ is lowered.

From Table 1, we see that since the combination
$5\Delta\tilde{b}_1 - 12\Delta\tilde{b}_2 + 7\Delta\tilde{b}_3$ 
can only change by $\pm 6$ for complete pairs of hypermultiplets we can
not
find any extra states that will unify the couplings as MSSM.
However, as shown in Table 3h, the extra hypermultiplet pair
$(2,3)_{\pm 5/6}$ has
$5\Delta\tilde{b}_1 - 12\Delta\tilde{b}_2 + 7\Delta\tilde{b}_3 = 6$
and can have acceptable $\alpha_3(M_Z)$ for $\mu_0$
down to intermediate scales.

Finally, one can also consider the case where the MSSM Higgs 
doublets have no Kaluza-Klein towers. In this case
the beta functions in (\ref{minddg}) change to
\begin{equation}
\tilde{b}^{min,0higgshm}_i=(0,-4,-6)+\eta(4,4,4)
\end{equation}
and the unification condition corresponding to (\ref{thecondn}) is
\begin{equation}
5\Delta\tilde{b}_1 - 12\Delta\tilde{b}_2 + 7\Delta\tilde{b}_3 = -6
\, \, .
\label{nohmcond}
\end{equation}
In contrast to the minimal case (\ref{minddg}) (see also Table 2)
here we note that with no extra states
the prediction for $\alpha_3(M_Z)$ is {\it lower} as $\mu_0$ goes down,
since now the no-extra-state condition has
$5\Delta\tilde{b}_1 - 12\Delta\tilde{b}_2 + 7\Delta\tilde{b}_3 = 0$
which is {\it higher} than condition (\ref{nohmcond}), in accordance
with the phenomenon described above.
We find for this case $\alpha_3(M_Z)=0.119$ around $\mu_0=10^{15}$GeV
as shown in Table 3i.
We see from Table 1, that, again, it is easy to find simple extensions
for which unification is guaranteed for all scales $\mu_0$.
For example Table 3j lists the predictions for $\alpha_3(M_Z)$ when
$(1,1)_{\pm 1}+(2,1)_{\pm 1/2}$ 
hypermultiplets are added.

\bigskip
\bigskip

{\it Nonsupersymmetric Unification.} \hspace{0.5cm} 
As we saw in the supersymmetric case, unification with extra
dimensions at lower scales requires extra states at
intermediate scales. Also one needs to get rid of the proton
decay either by a non-$SU(5)$ unification or some stringy argument.
Thus with the introduction of intermediate scales and not
taking into account limits from proton decay,
one can then ask why not do away with supersymmetry altogether
and study nonsupersymmetric unification.

In Ref. \cite{ddg98} it was actually pointed
out that nonsupersymmetric unification is indeed possible with
extra dimensions if one adds extra states at $\mu_0$. In this Section we
study this issue more systematically. We find that it is very easy to find
simple representations which unify the gauge couplings.

The procedure will be somewhat different than in the supersymmetric
case, since the Standard Model does not unify and one cannot use
equations (\ref{mssmcond}). Here we study unification only at 
one-loop.
Eliminating $\alpha^{-1}(\mu_0)$ we get
from (\ref{ddg1}) and (\ref{ddg2})
\begin{equation} 
\alpha^{-1}_i(M_Z) = \alpha^{-1}(\Lambda) +
 { {b_i} \over {2\pi}} 
{\rm ln} \left[ {\Lambda \over M_Z} \right]
-{ {\tilde{b}_i} \over {2\pi}}
\left[ 
{\rm ln} \left[ {\Lambda \over \mu_0} \right]
-{{X_\delta} \over {\delta}}\left(
\left({\Lambda \over \mu_0}\right)^\delta - 1 \right) 
\right], 
\label{nonsseqns}
\end{equation}
where we have assumed unification at scale $\Lambda$, and 
$b_i=(41/10,-19/6,-7)$ are the standard model beta functions.
Similarly to the supersymmetric case, we assume that the couplings unify
with addition of some extra matter at the scale $\mu_0$
\begin{equation}
\tilde{b}_i = \tilde{b}_{SMi}^{min} + \Delta\tilde{b}_i
\, \, ,
\end{equation}
where $\tilde{b}_{SMi}^{min} = (1/10,-41/6,-21/2)$.

The procedure now is as follows. From Table 4, we use the 
$\Delta\tilde{b}_i$ for all the $SU(3) \times SU(2) \times U(1)$ 
states (scalars or fermions) that are contained in $SU(5)$ 
mulitplets up to the representation $75$.

Then from the three equations (\ref{nonsseqns})
we can find the solutions for $\alpha^{-1}(\Lambda)$,
$\Lambda$ and $\Lambda/\mu_0$ in the form
\begin{equation}
\alpha^{-1}(\Lambda) = {
	{ \left|  {
		\begin{array}{ccc}
			\alpha^{-1}_1(M_Z) & b_1 & \tilde{b}_1 \\
			\alpha^{-1}_2(M_Z) & b_2 & \tilde{b}_2 \\
			\alpha^{-1}_3(M_Z) & b_3 & \tilde{b}_3
		 \end{array} 
	  }
	  \right|  } 
	\over {\rm det} }
\, \, ,
\label{eqn1}
\end{equation}
\begin{equation}
{1 \over {2\pi}}{\rm ln}{\Lambda \over M_Z}
	= {
	{ \left|  {
		\begin{array}{ccc}
			1 & \alpha^{-1}_1(M_Z) & \tilde{b}_1 \\
			1 & \alpha^{-1}_2(M_Z) & \tilde{b}_2 \\
			1 & \alpha^{-1}_3(M_Z) & \tilde{b}_3
		 \end{array} 
	  }
	  \right|  } 
	\over {\rm det} }
\, \, ,
\label{eqn2}
\end{equation}
\begin{equation}
-{1 \over {2\pi}} \left[ 
{\rm ln} \left[ {\Lambda \over \mu_0} \right]
-{{X_\delta} \over {\delta}}\left(
\left({\Lambda \over \mu_0}\right)^\delta - 1 \right) 
\right] = {
	{ \left|  {
		\begin{array}{ccc}
			1 & b_1 & \alpha^{-1}_1(M_Z) \\
			1 & b_2 & \alpha^{-1}_2(M_Z) \\
			1 & b_3 & \alpha^{-1}_3(M_Z) \\
		 \end{array} 
	  }
	  \right|  } 
	\over {\rm det} } \approx { 43.01 \over {\rm det}}
\, \, ,
\label{eqn3}
\end{equation}
where
\begin{equation}
{\rm det} = 
	\left|  {
		\begin{array}{ccc}
			1 & b_1 & \tilde{b}_1 \\
			1 & b_2 & \tilde{b}_2 \\
			1 & b_3 & \tilde{b}_3
		 \end{array} 
	  }
	  \right| = 
		-{23 \over 6} \Delta\tilde{b}_1
		+{111 \over 10} \Delta\tilde{b}_2
		-{109 \over 15} \Delta\tilde{b}_3 + {1 \over 15}
\, \, .
\end{equation}

The condition for perturbative unification is 
$N_{KK} \alpha^{-1}(\Lambda) > 1$,
while the requirements $1.0 {\rm TeV} < \Lambda < M_{Pl}$ and $\Lambda/\mu_0 >1$
give respectively $0.4 < \frac{1}{2\pi} ln \frac{\Lambda}{M_Z} < 6.2$ and det $> 0$.

In particular the inequality det $> 0$ tells us that we need a nontrivial
SU(2) quantum number, since in det the factor $1/15$ is too small
and only $\Delta\tilde{b}_2$ has a positive coefficient.
Indeed, if we look at the Table 4, in the last
column one can see that there is no unification for any SU(2) singlet
states or for most of the SU(2) doublet states. 
However, once the states do obey the above inequalities, one can always
find values for $\Lambda/\mu_0$ and $\delta$ for which unification takes
place.

In Table 5 we list the five states with lowest $\Lambda$.
The first case in Table 5 with one extra fermion
$(4, 1)_{3/2}$ has a $\mu_0$ which is too close to the
weak scale and is ruled out  empirically. However, the other states
which can have $\mu_0$ in tens of TeVs are perfectly allowable.
Other states listed in Table 4 that do have
unification at intermediate scales might be relevant for
indirect signals, like neutrino masses.
The highest $\Lambda$ found is for an extra fermion $(3,6)_{1/3}$ and 
it is $\Lambda=1.3 \times 10^{14}$GeV. Thus we see that there are many
possible minimal extensions of the nonsupersymmetric models for
which we have an intermediate unification.

\bigskip
\bigskip

{\it Conclusions.} \hspace{0.5cm} Lower mass extra dimensions offer the
prospect of experimental tests
such as
submillimeter modification of gravity, 
missing energy in collider experiments,
and departures from the Standard Model, in the
foreseeable future.

We have shown that with enlarged extra dimensions, unification of the
gauge couplings can be maintained in the supersymmetric case by including
certain extra states which have been identified by examining
systematically all of the $SU(3) \times SU(2) \times U(1)$
representations that are contained in $SU(5)$ irreducible representations
up to the {\bf 75}.  In fact, the unification can be even more accurate
than in the MSSM where the value of $\alpha_3(M_Z)$ is slightly high
($\alpha_3(M_Z)\simeq0.126$) compared to the data
($\alpha_3(M_Z)=0.119\pm0.002$). 

Similarly ``perfect'' unification has been demonstrated in the 
non-supersymmetric case. After all, since the gauge hierarchy
has been itself ameliorated by the enlargement of the extra
dimensions, the motivation for supersymmetry has been correspondingly
weakened. Thus, the non-SUSY possibility becomes of more interest.

In our opinion, the Achilles heel of this whole approach
lies in the subtle question of the proton stability. 
On one hand this could of course be cured by a suitable choice of
a unification group. On the other hand,
the mechanisms that avoid proton decay proposed in the
literature, while plausible, certainly merit further investigation for
their consistency.

With that one caveat, we feel that the idea of the enlarged
Kaluza-Klein dimensions is a very exciting one, particularly as
it inspires new insights into old theoretical prejudices.

\bigskip
\bigskip
\bigskip

{\bf Acknowledgments  } \hspace{0.5cm} 

We thank Z. Kakushadze for useful remarks.
This work was supported in part by the US Department of Energy
under the Grant No. DE-FG02-97ER-41036.

\bigskip
\bigskip

{\bf Note added:} After completion of this paper, an article\cite{dq99}
has been brought to our attention, which has some similarity to 
the part of our work in the supersymmetric case.

\newpage

\begin{center}
 \framebox{\begin{tabular}{l|l|l|l|l}
$SU(2) \times SU(3)_{Y/2}$ state 
& $\Delta\tilde{b}_1$ & $\Delta\tilde{b}_2$ & $\Delta\tilde{b}_3$
& $5\Delta\tilde{b}_1 - 12\Delta\tilde{b}_2 + 7\Delta\tilde{b}_3$
 \\ \hline &&&&\\ 
$(3,1)_1+(3,1)_{-1}$ 
&\hspace{0.1cm}36/5&\hspace{0.3cm}8 &\hspace{0.3cm}0 &\hspace{2cm}-60\\ 
$(3,3)_{4/3}+(3,\bar{3})_{-4/3}$
&192/5&\hspace{0.3cm}24 &\hspace{0.3cm}6 &
\hspace{2cm}-54\\
$(3,1)_0$
&\hspace{0.4cm}0&\hspace{0.3cm}4 &\hspace{0.3cm}0 & \hspace{2cm}-48\\
$(2,3)_{1/6}+(2,\bar{3})_{-1/6}$
&\hspace{0.2cm}2/5&\hspace{0.3cm}6 &\hspace{0.3cm}4 & \hspace{2cm}-42\\
$(2,1)_{1/2}+(2,1)_{-1/2}$
&\hspace{0.2cm}6/5&\hspace{0.3cm}2 &\hspace{0.3cm}0 & \hspace{2cm}-18\\
$(2,6)_{1/6}+(2,\bar{6})_{-1/6}$
&\hspace{0.2cm}4/5&\hspace{0.1cm}12 &\hspace{0.1cm}20 & \hspace{2.2cm}0\\
$(2,3)_{5/6}+(2,\bar{3})_{-5/6}$
&\hspace{0.3cm}10&\hspace{0.3cm}6 &\hspace{0.3cm}4 & \hspace{2.2cm}6\\
$(1,1)_{1}+(1,1)_{-1}$
&\hspace{0.1cm}12/5&\hspace{0.3cm}0 &\hspace{0.3cm}0 & \hspace{2.1cm}12\\
$(1,3)_{1/3}+(1,\bar{3})_{-1/3}$
&\hspace{0.2cm}4/5&\hspace{0.3cm}0 &\hspace{0.3cm}2 & \hspace{2.1cm}18\\
$(2,8)_{1/2}+(2,8)_{-1/2}$
&\hspace{0.1cm}48/5&\hspace{0.1cm}16 &\hspace{0.1cm}24 &
\hspace{2.1cm}24\\
$(1,3)_{2/3}+(1,\bar{3})_{-2/3}$
&\hspace{0.1cm}16/5&\hspace{0.3cm}0 &\hspace{0.3cm}2 & \hspace{2.1cm}30\\
$(2,1)_{3/2}+(2,1)_{-3/2}$
&\hspace{0.1cm}54/5&\hspace{0.3cm}2 &\hspace{0.3cm}0 & \hspace{2.1cm}30\\
$(1,8)_{0}$
&\hspace{0.4cm}0&\hspace{0.3cm}0 &\hspace{0.3cm}6 & \hspace{2.1cm}42\\
$(1,1)_{2}+(1,1)_{-2}$
&\hspace{0.1cm}48/5&\hspace{0.3cm}0 &\hspace{0.3cm}0 & \hspace{2.1cm}48\\
$(2,3)_{7/6}+(2,\bar{3})_{-7/6}$
&\hspace{0.1cm}98/5&\hspace{0.3cm}6 &\hspace{0.3cm}4 & \hspace{2.1cm}54\\
\end{tabular}}
\end{center}

\vspace{0.5cm}
{\bf Table 1}: Extra hypermultiplets for which the combination of 
beta functions $
5\Delta\tilde{b}_1 - 12\Delta\tilde{b}_2 + 7\Delta\tilde{b}_3$
is between -60 and 60.

\begin{center}
 \framebox{\begin{tabular}{l|l|l|l|l|l}
$\delta$ & $\rho$ & \hspace{1.0cm}$\Lambda$ & \hspace{1.0cm}$\mu_0$ &
$\alpha_3(M_Z)$ & $\alpha(\Lambda)$ \\
\hline
\hline
1 & 0.0 & 3.00e+16 & 3.00e+16 & 0.1260 & 0.0433\\
1 & 0.4 & 2.13e+16 & 1.43e+16 & 0.1284 & 0.0428\\
1 & 0.8 & 1.15e+16 & 5.16e+15 & 0.1331 & 0.0420\\
1 & 1.2 & 4.08e+15 & 1.23e+15 & 0.1420 & 0.0407\\
1 & 1.6 & 7.84e+14 & 1.58e+14 & 0.1592 & 0.0389\\
1 & 2.0 & 6.00e+13 & 8.13e+12 & 0.1970 & 0.0363\\
1 & 2.4 & 1.19e+14 & 1.09e+11 & 0.3142 & 0.0330\\
\hline
2 & 0.0 & 3.00e+16 & 3.00e+16 & 0.1260 & 0.0433\\
2 & 0.8 & 1.33e+15 & 5.97e+14 & 0.1538 & 0.0395\\
2 & 1.2 & 7.09e+12 & 2.14e+12 & 0.2492 & 0.0344\\
\hline
3 & 0.0 & 3.00e+16 & 3.00e+16 & 0.1260 & 0.0433\\
3 & 0.4 & 5.84e+15 & 3.91e+15 & 0.1392 & 0.0412\\
3 & 0.8 & 1.54e+13 & 6.93e+12 & 0.2286 & 0.0351\\
\end{tabular}}
\end{center}

\vspace{0.5cm}
{\bf Table 2}:
Predictions for $\alpha_3(M_z)$ and $\alpha(\Lambda)$
in the supersymmetric model with enlarged Kaluza-Klein
dimensions with two KK towers for MSSM Higgs and no extra
states. Inputs are the number of extra dimensions $\delta$ and 
$\rho\equiv{\rm ln}(\Lambda/\mu_0)$. Number of 
chiral generations with Kaluza-Klein towers has been
set to zero ($\eta=0$).

\newpage

\begin{center}
 \framebox{\begin{tabular}{l|l|l|l|l|l}
$\delta$ & $\rho$ & \hspace{1.0cm}$\Lambda$ & \hspace{1.0cm}$\mu_0$ &
$\alpha_3(M_Z)$ & $\alpha(\Lambda)$ \\
\hline
\hline
1 & 0.0 & 3.00e+16 & 3.00e+16 & 0.1260 & 0.0433\\
1 & 0.8 & 5.54e+15 & 2.49e+15 & 0.1250 & 0.0417\\
1 & 1.6 & 4.84e+13 & 9.78e+12 & 0.1228 & 0.0379\\
1 & 2.4 & 4.71e+08 & 4.27e+07 & 0.1179 & 0.0310\\
1 & 2.6 & 4.00e+06 & 2.97e+05 & 0.1158 & 0.0289\\
\hline
2 & 0.0 & 3.00e+16 & 3.00e+16 & 0.1260 & 0.0433\\
2 & 0.8 & 1.22e+14 & 5.48e+13 & 0.1235 & 0.0386\\
2 & 1.4 & 2.38e+06 & 5.86e+05 & 0.1161 & 0.0286\\
\hline
3 & 0.0 & 3.00e+16 & 3.00e+16 & 0.1260 & 0.0433\\
3 & 0.6 & 4.25e+13 & 2.33e+13 & 0.1231 & 0.0379\\
3 & 1.0 & 1.40e+05 & 5.15e+04 & 0.1151 & 0.0275\\
\end{tabular}}
\end{center}

\vspace{0.5cm}
{\bf Table 3a}:
Extra pair of $(1,1)_{\pm 1}$ multiplets above $\mu_0$: 
predictions for $\alpha_3(M_z)$ and $\alpha(\Lambda)$.
Inputs are the number of extra dimensions $\delta$ and 
$\rho\equiv{\rm ln}(\Lambda/\mu_0)$. Number of 
chiral generations with Kaluza-Klein towers has been
set to zero ($\eta=0$).
The extra states have 
$ 5\Delta\tilde{b}_1 - 12\Delta\tilde{b}_2 + 7\Delta\tilde{b}_3 =12$.

\begin{center}
 \framebox{\begin{tabular}{l|l|l|l|l|l}
$\delta$ & $\rho$ & \hspace{1.0cm}$\Lambda$ & \hspace{1.0cm}$\mu_0$ &
$\alpha_3(M_Z)$ & $\alpha(\Lambda)$ \\
\hline
\hline
1 & 0.0 & 3.00e+16 & 3.00e+16 & 0.1260 & 0.0433\\
1 & 1.2 & 8.96e+14 & 2.70e+14 & 0.1241 & 0.0420\\
1 & 2.2 & 2.25e+10 & 2.49e+09 & 0.1195 & 0.0387\\
1 & 2.6 & 4.00e+06 & 2.97e+05 & 0.1158 & 0.0362\\
\hline
2 & 0.0 & 3.00e+16 & 3.00e+16 & 0.1260 & 0.0433\\
2 & 0.8 & 1.22e+14 & 5.48e+13 & 0.1235 & 0.0414\\
2 & 1.2 & 1.13e+10 & 3.39e+09 & 0.1196 & 0.0385\\
2 & 1.4 & 2.38e+06 & 5.86e+05 & 0.1161 & 0.0362\\
\hline
3 & 0.0 & 3.00e+16 & 3.00e+16 & 0.1260 & 0.0433\\
3 & 0.8 & 4.47e+10 & 2.01e+10 & 0.1203 & 0.0389\\
3 & 1.0 & 1.40e+05 & 5.15e+04 & 0.1151 & 0.0355\\
\end{tabular}}
\end{center}

\vspace{0.5cm}
{\bf Table 3b}: 
Extra pairs of $(1,3)_{\pm 2/3}+(2,1)_{\pm 1/2}$ hypermultiplets
above $\mu_0$:
predictions for $\alpha_3(M_z)$ and $\alpha(\Lambda)$.
Inputs are the number of extra dimensions $\delta$ and 
$\rho\equiv{\rm ln}(\Lambda/\mu_0)$. Number of 
chiral generations with Kaluza-Klein towers has been
set to zero ($\eta=0$).
The extra states have 
$ 5\Delta\tilde{b}_1 - 12\Delta\tilde{b}_2 + 7\Delta\tilde{b}_3 =12$.

\newpage

\begin{center}
 \framebox{\begin{tabular}{l|l|l|l|l|l}
$\delta$ & $\rho$ & \hspace{1.0cm}$\Lambda$ & \hspace{1.0cm}$\mu_0$ &
$\alpha_3(M_Z)$ & $\alpha(\Lambda)$ \\
\hline
\hline
1 & 0.0 & 3.00e+16 & 3.00e+16 & 0.1260 & 0.0433\\
1 & 0.8 & 1.04e+15 & 4.66e+14 & 0.1243 & 0.0430\\
1 & 1.2 & 2.72e+13 & 8.19e+12 & 0.1227 & 0.0428\\
1 & 1.6 & 7.98e+10 & 1.61e+10 & 0.1202 & 0.0424\\
1 & 1.8 & 1.38e+09 & 2.28e+08 & 0.1185 & 0.0421\\
1 & 2.0 & 8.87e+06 & 1.20e+06 & 0.1164 & 0.0417\\
\hline
2 & 0.0 & 3.00e+16 & 3.00e+16 & 0.1260 & 0.0433\\
2 & 0.4 & 1.35e+15 & 9.03e+14 & 0.1246 & 0.0431\\
2 & 1.0 & 3.13e+08 & 1.15e+08 & 0.1183 & 0.0420\\
\hline
3 & 0.0 & 3.00e+16 & 3.00e+16 & 0.1260 & 0.0433\\
3 & 0.4 & 9.34e+13 & 6.26e+13 & 0.1235 & 0.0429\\
3 & 0.6 & 6.05e+10 & 3.32e+10 & 0.1205 & 0.0424\\
\end{tabular}}
\end{center}

\vspace{0.5cm}
{\bf Table 3c}: 
Extra pairs of $(2,1)_{\pm 3/2}+(2,1)_{\pm 1/2}$ hypermultiplets
above $\mu_0$: 
predictions for $\alpha_3(M_z)$ and $\alpha(\Lambda)$.
Inputs are the number of extra dimensions $\delta$ and 
$\rho\equiv{\rm ln}(\Lambda/\mu_0)$. Number of 
chiral generations with Kaluza-Klein towers has been
set to zero ($\eta=0$).
The extra states have 
$ 5\Delta\tilde{b}_1 - 12\Delta\tilde{b}_2 + 7\Delta\tilde{b}_3 = 12$.

\begin{center}
 \framebox{\begin{tabular}{l|l|l|l|l|l}
$\delta$ & $\rho$ & \hspace{1.0cm}$\Lambda$ & \hspace{1.0cm}$\mu_0$ &
$\alpha_3(M_Z)$ & $\alpha(\Lambda)$ \\
\hline
\hline
1 & 0.0 & 3.00e+16 & 3.00e+16 & 0.1260 & 0.0433\\
1 & 0.4 & 9.12e+15 & 6.11e+15 & 0.1254 & 0.0446\\
1 & 0.8 & 1.04e+15 & 4.66e+14 & 0.1243 & 0.0473\\
1 & 1.2 & 2.72e+13 & 8.19e+12 & 0.1227 & 0.0526\\
1 & 1.6 & 7.98e+12 & 1.61e+10 & 0.1202 & 0.0642\\
1 & 2.0 & 8.87e+06 & 1.20e+06 & 0.1164 & 0.0976\\
\hline
2 & 0.0 & 3.00e+16 & 3.00e+16 & 0.1260 & 0.0433\\
2 & 0.6 & 6.14e+13 & 3.37e+13 & 0.1233 & 0.0514\\
2 & 1.0 & 3.13e+08 & 1.15e+08 & 0.1183 & 0.0814\\
\hline
3 & 0.0 & 3.00e+16 & 3.00e+16 & 0.1260 & 0.0433\\
3 & 0.2 & 4.37e+15 & 3.57e+15 & 0.1251 & 0.0455\\
3 & 0.6 & 6.05e+10 & 3.32e+10 & 0.1205 & 0.0651\\
\end{tabular}}
\end{center}

\vspace{0.5cm}
{\bf Table 3d}: 
Two extra pairs of $(2,3)_{\pm 5/6}$ hypermultiplets
(or extra pairs of $(2,3)_{\pm 7/6}+(2,3)_{\pm 1/6}$ hypermultiplets
; they have the same $\Delta\tilde{b}$s)
above $\mu_0$:
predictions for $\alpha_3(M_z)$ and $\alpha(\Lambda)$.
Inputs are the number of extra dimensions $\delta$ and 
$\rho\equiv{\rm ln}(\Lambda/\mu_0)$. Number of 
chiral generations with Kaluza-Klein towers has been
set to zero ($\eta=0$).
The extra states have 
$ 5\Delta\tilde{b}_1 - 12\Delta\tilde{b}_2 + 7\Delta\tilde{b}_3 = 12$.

\newpage

\begin{center}
 \framebox{\begin{tabular}{l|l|l|l|l|l}
$\delta$ & $\rho$ & \hspace{1.0cm}$\Lambda$ & \hspace{1.0cm}$\mu_0$ &
$\alpha_3(M_Z)$ & $\alpha(\Lambda)$ \\
\hline
\hline
1 & 0.0 & 3.00e+16 & 3.00e+16 & 0.1260 & 0.0433\\
1 & 1.0 & 5.02e+15 & 1.85e+15 & 0.1197 & 0.0412\\
1 & 1.2 & 2.41e+15 & 7.26e+14 & 0.1174 & 0.0404\\
1 & 1.6 & 2.98e+14 & 6.01e+13 & 0.1113 & 0.0384\\
\hline
2 & 0.0 & 3.00e+16 & 3.00e+16 & 0.1260 & 0.0433\\
2 & 0.4 & 9.83e+15 & 6.59e+15 & 0.1221 & 0.0420\\
2 & 0.8 & 5.79e+14 & 2.60e+14 & 0.1136 & 0.0390\\
\hline
3 & 0.0 & 3.00e+16 & 3.00e+16 & 0.1260 & 0.0433\\
3 & 0.4 & 3.78e+15 & 2.53e+15 & 0.1190 & 0.0409\\
3 & 0.6 & 2.72e+14 & 1.49e+14 & 0.1113 & 0.0383\\
\end{tabular}}
\end{center}

\vspace{0.5cm}
{\bf Table 3e}: 
Extra pair of $(1,3)_{\pm 1/3}$ hypermultiplets
above $\mu_0$: 
predictions for $\alpha_3(M_z)$ and $\alpha(\Lambda)$.
Inputs are the number of extra dimensions $\delta$ and 
$\rho\equiv{\rm ln}(\Lambda/\mu_0)$. Number of 
chiral generations with Kaluza-Klein towers has been
set to zero ($\eta=0$).
The extra states have 
$ 5\Delta\tilde{b}_1 - 12\Delta\tilde{b}_2 + 7\Delta\tilde{b}_3 =18$.

\begin{center}
 \framebox{\begin{tabular}{l|l|l|l|l|l}
$\delta$ & $\rho$ & \hspace{1.0cm}$\Lambda$ & \hspace{1.0cm}$\mu_0$ &
$\alpha_3(M_Z)$ & $\alpha(\Lambda)$ \\
\hline
\hline
1 & 0.0 & 3.00e+16 & 3.00e+16 & 0.1260 & 0.0433\\
1 & 1.0 & 4.23e+14 & 1.55e+14 & 0.1188 & 0.0446\\
1 & 1.2 & 7.32e+13 & 2.21e+13 & 0.1162 & 0.0452\\
1 & 1.6 & 4.92e+11 & 9.94e+13 & 0.1093 & 0.0470\\
\hline
2 & 0.0 & 3.00e+16 & 3.00e+16 & 0.1260 & 0.0433\\
2 & 0.4 & 2.09e+15 & 1.40e+15 & 0.1215 & 0.0441\\
2 & 0.8 & 2.38e+12 & 1.07e+12 & 0.1116 & 0.0465\\
\hline
3 & 0.0 & 3.00e+16 & 3.00e+16 & 0.1260 & 0.0433\\
3 & 0.4 & 2.11e+14 & 1.42e+14 & 0.1180 & 0.0449\\
3 & 0.6 & 3.89e+11 & 2.13e+11 & 0.1093 & 0.0472\\
\end{tabular}}
\end{center}

\vspace{0.5cm}
{\bf Table 3f}: 
Extra pairs of $(1,1)_{\pm 1}+(2,3)_{\pm 5/6}$ hypermultiplets
above $\mu_0$: 
predictions for $\alpha_3(M_z)$ and $\alpha(\Lambda)$.
Inputs are the number of extra dimensions $\delta$ and 
$\rho\equiv{\rm ln}(\Lambda/\mu_0)$. Number of 
chiral generations with Kaluza-Klein towers has been
set to zero ($\eta=0$).
The extra states have 
$ 5\Delta\tilde{b}_1 - 12\Delta\tilde{b}_2 + 7\Delta\tilde{b}_3 =18$.

\newpage

\begin{center}
 \framebox{\begin{tabular}{l|l|l|l|l|l}
$\delta$ & $\rho$ & \hspace{1.0cm}$\Lambda$ & \hspace{1.0cm}$\mu_0$ &
$\alpha_3(M_Z)$ & $\alpha(\Lambda)$ \\
\hline
\hline
1 & 0.0 & 3.00e+16 & 3.00e+16 & 0.1260 & 0.0433\\
1 & 1.0 & 1.24e+15 & 4.55e+14 & 0.1299 & 0.0451\\
1 & 2.0 & 2.29e+10 & 3.10e+09 & 0.1460 & 0.0528\\
\hline
2 & 0.0 & 3.00e+16 & 3.00e+16 & 0.1260 & 0.0433\\
2 & 0.8 & 2.57e+13 & 1.16e+13 & 0.1356 & 0.0477\\
2 & 1.2 & 1.69e+08 & 5.10e+07 & 0.1552 & 0.0573\\
\hline
3 & 0.0 & 3.00e+16 & 3.00e+16 & 0.1260 & 0.0433\\
3 & 0.4 & 7.38e+14 & 4.95e+14 & 0.1309 & 0.0455\\
3 & 0.6 & 6.64e+12 & 3.65e+12 & 0.1377 & 0.0486\\
\end{tabular}}
\end{center}

\vspace{0.5cm}
{\bf Table 3g}: 
Extra pair of $(2,3)_{\pm 5/6}$ hypermultiplets
above $\mu_0$: 
predictions for $\alpha_3(M_z)$ and $\alpha(\Lambda)$.
Inputs are the number of extra dimensions $\delta$ and 
$\rho\equiv{\rm ln}(\Lambda/\mu_0)$. Number of 
chiral generations with Kaluza-Klein towers has been
set to zero ($\eta=0$).
The extra states have 
$ 5\Delta\tilde{b}_1 - 12\Delta\tilde{b}_2 + 7\Delta\tilde{b}_3 =6$.

\begin{center}
 \framebox{\begin{tabular}{l|l|l|l|l|l}
$\delta$ & $\rho$ & \hspace{1.0cm}$\Lambda$ & \hspace{1.0cm}$\mu_0$ &
$\alpha_3(M_Z)$ & $\alpha(\Lambda)$
 \\ \hline \hline
1 & 0.0 & 3.00e+16 & 3.00e+16 & 0.1260 & 0.0433\\
1 & 0.6 & 7.04e+15 & 3.87e+15 & 0.1240 & 0.0440\\
1 & 1.0 & 1.03e+15 & 3.77e+14 & 0.1216 & 0.0442\\
1 & 1.4 & 4.43e+13 & 1.09e+13 & 0.1180 & 0.0451\\
1 & 1.8 & 3.11e+04 & 5.14e+10 & 0.1128 & 0.0467\\ 
\hline
2 & 0.0 & 3.00e+16 & 3.00e+16 & 0.1260 & 0.0433\\
2 & 0.4 & 3.64e+15 & 2.44e+15 & 0.1233 & 0.0439\\
2 & 0.8 & 1.70e+13 & 7.65e+12 & 0.1172 & 0.0455\\
2 & 1.0 & 1.14e+11 & 4.19e+10 & 0.1121 & 0.0471\\ 
\hline
3 & 0.0 & 3.00e+16 & 3.00e+16 & 0.1260 & 0.0433\\
3 & 0.4 & 5.94e+14 & 3.98e+14 & 0.1212 & 0.0444\\
3 & 0.6 & 4.06e+12 & 2.23e+12 & 0.1158 & 0.0460\\
\end{tabular}}
\end{center}

\vspace{0.5cm}
{\bf Table 3h}: 
Case with only one Higgs hypermultiplet and
extra pair of $(2,3)_{\pm 5/6}$ hypermultiplets 
above $\mu_0$: 
predictions for $\alpha_3(M_z)$ and $\alpha(\Lambda)$.
Inputs are the number of extra dimensions $\delta$ and 
$\rho\equiv{\rm ln}(\Lambda/\mu_0)$. Number of 
chiral generations with Kaluza-Klein towers has been
set to zero ($\eta=0$).
The extra states have 
$ 5\Delta\tilde{b}_1 - 12\Delta\tilde{b}_2 + 7\Delta\tilde{b}_3 = 6$.

\newpage

\begin{center}
 \framebox{\begin{tabular}{l|l|l|l|l|l}
$\delta$ & $\rho$ & \hspace{1.0cm}$\Lambda$ & \hspace{1.0cm}$\mu_0$ &
$\alpha_3(M_Z)$ & $\alpha(\Lambda)$ \\
\hline
\hline
1 & 0.0 & 3.00e+16 & 3.00e+16 & 0.1260 & 0.0433\\
1 & 0.6 & 1.39e+16 & 7.63e+15 & 0.1232 & 0.0418\\
1 & 1.0 & 5.02e+15 & 1.85e+15 & 0.1197 & 0.0399\\
1 & 1.4 & 9.53e+14 & 2.35e+14 & 0.1146 & 0.0372\\
1 & 1.6 & 2.98e+14 & 6.01e+13 & 0.1113 & 0.0356\\
\hline
2 & 0.0 & 3.00e+16 & 3.00e+16 & 0.1260 & 0.0433\\
2 & 0.4 & 9.83e+15 & 6.59e+15 & 0.1221 & 0.0411\\
2 & 0.8 & 5.79e+14 & 2.60e+14 & 0.1134 & 0.0366\\
\hline
3 & 0.0 & 3.00e+16 & 3.00e+16 & 0.1260 & 0.0433\\
3 & 0.4 & 3.78e+15 & 2.53e+15 & 0.1190 & 0.0395\\
3 & 0.6 & 2.72e+14 & 1.49e+14 & 0.1113 & 0.0355\\
\end{tabular}}
\end{center}

\vspace{0.5cm}
{\bf Table 3i}: 
Case with no Higgs hypermultiplets and
no extra hypermultiplets
above $\mu_0$: 
predictions for $\alpha_3(M_z)$ and $\alpha(\Lambda)$.
Inputs are the number of extra dimensions $\delta$ and 
$\rho\equiv{\rm ln}(\Lambda/\mu_0)$. Number of 
chiral generations with Kaluza-Klein towers has been
set to zero ($\eta=0$).

\begin{center}
 \framebox{\begin{tabular}{l|l|l|l|l|l}
$\delta$ & $\rho$ & \hspace{1.0cm}$\Lambda$ & \hspace{1.0cm}$\mu_0$ &
$\alpha_3(M_Z)$ & $\alpha(\Lambda)$ \\
\hline
\hline
1 & 0.0 & 3.00e+16 & 3.00e+16 & 0.1260 & 0.0433\\
1 & 0.6 & 1.03e+16 & 5.65e+15 & 0.1253 & 0.0423\\
1 & 1.0 & 2.49e+15 & 9.16e+14 & 0.1246 & 0.0410\\
1 & 1.4 & 2.46e+14 & 6.06e+13 & 0.1235 & 0.0391\\
1 & 1.8 & 6.37e+12 & 1.05e+12 & 0.1219 & 0.0365\\
1 & 2.2 & 2.25e+10 & 2.49e+09 & 0.1195 & 0.0330\\
1 & 2.6 & 4.00e+06 & 2.97e+05 & 0.1158 & 0.0288\\
\hline
2 & 0.0 & 3.00e+16 & 3.00e+16 & 0.1260 & 0.0433\\
2 & 0.4 & 6.34e+15 & 4.25e+15 & 0.1252 & 0.0419\\
2 & 0.8 & 1.22e+14 & 5.48e+13 & 0.1235 & 0.0386\\
2 & 1.2 & 1.13e+10 & 3.39e+09 & 0.1196 & 0.0327\\
2 & 1.4 & 2.38e+06 & 5.86e+05 & 0.1161 & 0.0286\\
\hline
3 & 0.0 & 3.00e+16 & 3.00e+16 & 0.1260 & 0.0433\\
3 & 0.6 & 4.25e+13 & 2.33e+13 & 0.1231 & 0.0379\\
3 & 0.8 & 4.47e+10 & 2.01e+10 & 0.1203 & 0.0335\\
3 & 1.0 & 1.40e+05 & 5.15e+04 & 0.1151 & 0.0275\\
\end{tabular}}
\end{center}

\vspace{0.5cm}
{\bf Table 3j}: 
Case with no Higgs hypermultiplets and
extra pairs of $(1,1)_{\pm 1}+(2,1)_{\pm 1/2}$ hypermultiplets
above $\mu_0$: 
predictions for $\alpha_3(M_z)$ and $\alpha(\Lambda)$.
Inputs are the number of extra dimensions $\delta$ and 
$\rho\equiv{\rm ln}(\Lambda/\mu_0)$. Number of 
chiral generations with Kaluza-Klein towers has been
set to zero ($\eta=0$).
The extra states have 
$ 5\Delta\tilde{b}_1 - 12\Delta\tilde{b}_2 + 7\Delta\tilde{b}_3 = -6$.


\begin{center}
 \framebox{\begin{tabular}{l|l|l|l|l} 
$SU(2) \times SU(3)_{Y/2}$ &
$\Delta\tilde{b}_1$ & 
$\Delta\tilde{b}_2$ & 
$\Delta\tilde{b}_3$ &  \hspace{0.3cm}
${1 \over {2\pi}}{\rm ln}(\Lambda/M_Z)$\\ 
\hline 
$(1,1)_{1}$
&\hspace{0.3cm}1/5&\hspace{0.3cm}0 &\hspace{0.3cm}0& 
\\ $(1,1)_{2}$
&\hspace{0.3cm}4/5&\hspace{0.3cm}0 &\hspace{0.3cm}0& 
\\ $(1,3)_{-1/3}$
&\hspace{0.3cm}1/15&\hspace{0.3cm}0 &\hspace{0.3cm}1/6&
\\ $(1,3)_{2/3}$
&\hspace{0.3cm}4/15&\hspace{0.3cm}0 &\hspace{0.3cm}1/6& 
\\ $(1,3)_{-4/3}$
&\hspace{0.3cm}16/15&\hspace{0.3cm}0 &\hspace{0.3cm}1/6& 
\\ $(1,3)_{5/3}$
&\hspace{0.3cm}5/3&\hspace{0.3cm}0 &\hspace{0.3cm}1/6& 
\\ $(1,6)_{1/3}$
&\hspace{0.3cm}2/15&\hspace{0.3cm}0 &\hspace{0.3cm}5/6& 
\\ $(1,6)_{-2/3}$
&\hspace{0.3cm}8/15&\hspace{0.3cm}0 &\hspace{0.3cm}5/6& 
\\ $(1,6)_{-4/3}$
&\hspace{0.3cm}32/15&\hspace{0.3cm}0 &\hspace{0.3cm}5/6& 
\\ $(1,8)_{0}$
&\hspace{0.3cm}0&\hspace{0.3cm}0 &\hspace{0.3cm}1/2&
\\ $(1,8)_{1}$
&\hspace{0.3cm}8/5&\hspace{0.3cm}0 &\hspace{0.3cm}1/2& 
\\ $(1,10)_{1}$
&\hspace{0.3cm}2&\hspace{0.3cm}0 &\hspace{0.3cm}5/2&
\\ $(1,15)_{-1/3}$
&\hspace{0.3cm}1/3&\hspace{0.3cm}0 &\hspace{0.3cm}10/3& 
\\$(1,15')_{-4/3}$&
\hspace{0.3cm}16/3&\hspace{0.3cm}0 &\hspace{0.3cm}35/6&
\\$(2,1)_{1/2}$
&1/10&\hspace{0.3cm}1/6 &\hspace{0.3cm}0&
\\ $(2,1)_{-3/2}$
&9/10&\hspace{0.3cm}1/6 &\hspace{0.3cm}0&
\\$(2,3)_{1/6}$
&\hspace{0.3cm}1/30&\hspace{0.3cm}1/2 &\hspace{0.3cm}1/3&F(1.53)
\\ $(2,3)_{5/6}$
&\hspace{0.3cm}5/6&\hspace{0.3cm}1/2 &\hspace{0.3cm}1/3&
\\ $(2,3)_{7/6}$
&\hspace{0.3cm}49/30&\hspace{0.3cm}1/2 &\hspace{0.3cm}1/3&
\\ $(2,6)_{1/6}$
&\hspace{0.3cm}1/15&\hspace{0.3cm}1 &\hspace{0.3cm}5/3&
\\ $(2,6)_{5/6}$
&\hspace{0.3cm}5/3&\hspace{0.3cm}1 &\hspace{0.3cm}5/3&
\\ $(2,6)_{-7/6}$
&\hspace{0.3cm}49/15&\hspace{0.3cm}1 &\hspace{0.3cm}5/3&
\\ $(2,8)_{-1/2}$
&\hspace{0.3cm}4/5&\hspace{0.3cm}4/3 &\hspace{0.3cm}2&
\\ $(2,10)_{-1/2}$
&\hspace{0.3cm}1&\hspace{0.3cm}5/3 &\hspace{0.3cm}5&
\\ $(3,1)_{1}$
&\hspace{0.3cm}3/5&\hspace{0.3cm}2/3 &\hspace{0.3cm}0&F(2.09)
\\ $(3,1)_{0}$
&\hspace{0.3cm}0&\hspace{0.3cm}1/3 &\hspace{0.3cm}0&F(1.79)
\\ $(3,3)_{-1/3}$
&\hspace{0.3cm}1/5&\hspace{0.3cm}2 &\hspace{0.3cm}1/2 &
S(2.32) {\rm or} F(4.04) 
\\$(3,3)_{2/3}$ 
&\hspace{0.3cm}4/5&\hspace{0.3cm}2 &\hspace{0.3cm}1/2&
S(1.84) {\rm or} F(3.81)
\\ $(3,3)_{-4/3}$ 
&\hspace{0.3cm}16/5&\hspace{0.3cm}2 &\hspace{0.3cm}1/2&F(1.27) 
\\ $(3,6)_{1/3}$ 
&\hspace{0.3cm}2/5&\hspace{0.3cm}4 &\hspace{0.3cm}5/2&
S(3.22) {\rm or} F(4.46)
\\$(3,8)_{0}$
&\hspace{0.3cm}0&\hspace{0.3cm}8/3 &\hspace{0.3cm}3/2&
S(2.67) {\rm or} F(4.31)
\\ $(4,1)_{1/2}$ 
&\hspace{0.3cm}1/5&\hspace{0.3cm}5/3 &\hspace{0.3cm}0&
S(2.20) {\rm or} F(3.93)
\\ $(4,1)_{3/2}$
&\hspace{0.3cm}9/5&\hspace{0.3cm}5/3 &\hspace{0.3cm}0&
S(0.43) {\rm or} F(3.07)
\\ $(4,3)_{7/6}$ 
&\hspace{0.3cm}49/15&\hspace{0.3cm}5 &\hspace{0.3cm}2/3& 
S(3.21) {\rm or} F(4.02)
\\ $(5,1)_{2}$
&\hspace{0.3cm}4&\hspace{0.3cm}10/3 &\hspace{0.3cm}0& 
S(1.95) {\rm or} F(3.36)
\\
\end{tabular}}

\end{center}

{\bf Table 4}: Beta functions for extra scalar 
$SU(3) \times SU(2) \times U(1)$ states that can
be found
in broken multiplets of $SU(5)$ up to the {\bf 75}. All scalars are
complex 
except $(3,1)_0$, $(1,8)_0$ and $(3,8)_0$. Beta functions for
corresponding
fermion pairs (Dirac fermions) are obtained by multiplying the above
scalar functions by 4. Supersymmetric beta functions for supermultiplets
with the same quantum numbers are obtained by multiplying the scalar beta
functions by 6 ({\it cf.} Table 1).
The last column indicates values for
${1 \over {2\pi}}{\rm ln}(\Lambda/M_Z)$ 
where nonsupersymmetric unification does occur, either with one
extra scalar (S) or 
one fermion pair (F) at $\mu_0$.

\newpage

\begin{center}
 \framebox{\begin{tabular}{l|c|c|l|l} 
Extra state & $\alpha^{-1}(\Lambda)$ & $\Lambda$ & $\delta$ &  $\mu_0$\\ 
\hline 
&&&&\\ 
S $(4,1)_{3/2}$ & $50.1$ & $1.4 \times 10^{3}GeV$ & 1&$1.0
\times 10^{2}GeV$\\
&&&$2$ & $3.4 \times 10^{2}GeV$\\
&&& $3$ & $5.4 \times 10^{2}GeV$\\
F $(3,3)_{-4/3}$ & $31.8$ & $2.7 \times 10^{5} GeV$&
$1$ & $3.6 \times 10^{4}GeV$\\
&&&$2$&$9.3 \times 10^{4}GeV$\\
&&&$3$&$1.3 \times 10^{5}GeV$\\
F $(2,3)_{1/6}$&$51.8$&$1.4 \times 10^{6} GeV$&
$1$&$1.1 \times 10^{5}GeV$\\
&&&$2$&$3.5 \times 10^{5}GeV$\\
&&&$3$&$5.4 \times 10^{5}GeV$\\
F $(3,1)_{0}$&$51.3$&$7.0 \times  10^{6} GeV$&
$1$&$6.2 \times 10^{5}GeV$\\
&&&$2$&$1.9 \times 10^{6}GeV$\\
&&&$3$&$2.9 \times 10^{6}GeV$\\

S $(3,3)_{2/3}$&$48.9$&
$9.6 \times 10^{6} GeV$&
$1$&$8.8 \times 10^{5}GeV$\\
&&&$2$&$2.7 \times 10^{6}GeV$\\
&&&$3$&$4.0 \times 10^{6}GeV$\\
\end{tabular}}

\end{center}
\vspace{0.5cm}
{\bf Table 5}: States at $\mu_0$ for which couplings unify in
extra dimensions.

\end{document}